\documentclass[]{article}
\usepackage{amsmath}
\usepackage{latexsym}
\usepackage{geometry}
\usepackage{hyperref}
\usepackage{breqn}
\hypersetup{
	colorlinks   = true,
	linkcolor    = blue,
	citecolor    = red
}
\usepackage{mathrsfs}
\title{
Emergence of cosmic space with Barrow entropy, in non-equilibrium thermodynamic conditions}

\author{Nandhida Krishnan.P\footnote{nandhidakrishnan@cusat.ac.in} , Titus K Mathew\footnote{titus@cusat.ac.in}
	\\$^{*\dagger}$Department of Physics, Cochin University of Science and Technology,\\ Kochi 682-022, India.\\$^\dagger$Inter University Center for Kerala Legacy in Astronomy and Mathematics.\\ $^\dagger$Centre for Particle Physics, CUSAT}

\begin{document}

\maketitle

\begin{abstract}
Recently, Barrow accounts for the quantum gravitational effects to the black hole surface. Thus the conventional area-entropy relation has modified, $S=(A/A_{0})^{1+\Delta/2},$ with an exponent $\Delta$, ranges $0\le\Delta\le1$, quantifies the amount of quantum gravitational deformation effect to the black hole surface. In recent literature, this horizon entropy has been extended to the cosmological context. Following this, we consider an n+1 dimensional non-flat universe with an apparent horizon as the boundary with appropriate temperature and associated entropy is Barrow entropy. We derived the modified form of the law of emergence from the equilibrium and non-equilibrium thermodynamic principles. Later studied the entropy maximization condition due to the modified law of emergence. On distinguishing the obtained result, it speculates that in order to hold the energy-momentum conservation, the universe with Barrow entropy as the horizon entropy should have non-equilibrium behaviour with an additional entropy production. However, the additional entropy production rate decreases over time, so the system eventually approaches equilibrium. Because of this, the constraint relation for entropy maximization looks similar for both equilibrium and non-equilibrium approaches. 
\end{abstract}
\section{Introduction}

The theory of black hole mechanics reveals the remarkable connection between gravity and thermodynamics. Bekenstein has conjectured that black hole exhibits entropy, which is proportional to the area of their horizon \cite{ PhysRevD.7.2333}. Meantime, adopting a semi-classical approach, Hawking has shown that black holes can be endowed with a temperature proportional to their surface gravity and behave like a thermal object, which emits radiation \cite{PhysRevD.13.191,Hawking1976}. The thermodynamic properties of the horizon of highly gravitating objects like black holes have become much interest \cite{Bardeen1973,Bekenstein1972,PhysRevD.9.3292,Hayward_1998,bekenstein1994we}. However, the thermodynamic properties of the horizon are not a mere unique property of black hole spacetime but are more general. Davies and Unruh found that a uniformly accelerating observer in the Minkowski spacetime can attribute a temperature $\mathcal{T}=a/2\pi,$ to the resulting horizon, where $a$ is the acceleration of the observer \cite{PCWDavies_1975,PhysRevD.14.870}. 
Later Jacobson \cite{PhysRevLett.75.1260} derived the Einstein's field equation from Clausius relation $\delta Q=\mathcal{T}\delta S$ together with the area-entropy relation for local Rindler observer. Here $\mathcal{T}$ is the Unruh temperature \cite{PhysRevD.14.870}, and $\delta Q$ is energy flux crossing the horizon, perceived by an accelerated observer just inside the horizon.  
Subsequently, Padmanabhan showed that it is possible to express the Einstein field equation as the thermodynamics relation $TdS=dE+PdV$ for a spherically symmetric horizon, where $P$ as the radial pressure and $T$ is the dynamical temperature of the horizon\cite{Padmanabhan_2002,PADMANABHAN200549}. Later Paranjape et al. extended this result to Gauss-Bonnet and more general Lancos-Lovelock gravity theories \cite{PhysRevD.74.104015}. Inspired by this,  Gibbons and Hawking \cite{PhysRevD.15.2738} argued that there is a vivid possibility for extending this analogy to the cosmological context for a local Rindler horizon. Such that, the cosmological event horizon in de Sitter space can be associated with a temperature, $1/2\pi r$, where $r$ is the radius of the cosmological horizon and entropy as given by the Bekenstein formula, $S=A/(4G)$. In the same spirit, Cai and Kim\cite{Cai_2005} derived the Friedmann equations from the Clausius relation by assuming Bekenstein entropy for the horizon and temperature as $\mathcal{T}=1/(2\pi \tilde{r}_{_A}),$ in the context of Einstein's gravity. Here $\tilde{r}_{_A}$ is the radius of the apparent horizon \cite{Cai_2009}. This has been extended to the Gauss-Bonnet and Lovelock gravity theories using the respective entropy relations. Later, the Friedmann equation has been obtained from the unified first law $dE=TdS+WdV$ proposed by Hayward \cite{Hayward_1998,HAYWARD1999347,PhysRevD.49.6467}, in the context of Einstein, Gauss-Bonnet and Lovelock gravity theories \cite{PhysRevD.75.084003}. Here $E$ is the Misner-Sharp energy inside the apparent horizon of volume $V$, the temperature is given by $T=\kappa/(2\pi)$($\kappa$ is the surface gravity)\cite{PhysRev.136.B571}, and $W=(\rho-P)/2,$ is the work density, in which $\rho$ and $P$ are the total density and pressure of the matter inside the horizon, respectively.\\

All the above works have indicates the intriguing connection between gravity and thermodynamics, which motivates more exciting speculations of the quantum nature of gravity, that one can interpret gravity as an emerging phenomenon with metric, curvature, etc., as the macroscopic variables which are being emerged from some underlying degrees of freedom. By trailing this notion, Verlinde \cite{Verlinde2011} interpreted gravity as an entropic force caused by the information change associated with the positions of material bodies in the universe. In this regard, the author successfully derived Newton's gravitational equation. Meantime, Padmanabhan arrived at the same result, through a different root, by employing the equipartition law of energy for the horizon degrees of freedom along with the thermodynamic relation connecting entropy S and active gravitational mass E, $S= E/2T$ \cite{doi:10.1142/S021773231003313X}. Here E
is the Komar energy, the source of gravitational acceleration, the temperature T determined from the surface gravity at the horizon, and the pressure P is provided by the source of Einstein's equations\cite{padmanabhan2004entropy}. Following the emergent gravity paradigm, Padmanabhan speculated that spacetime itself could be of emerging nature. However, it is challenging to consider time being emerged from some underlying degrees of freedom. However, this difficulty will be erased in cosmology due to the existence of a cosmic time for all inertial observers. This, in turn, motivates to propose that the cosmic space can be considered as emerging as cosmic time progresses \cite{padmanabhan2012emergence}. This notion leads to the proposal of the law of emergence, which explains that the expansion of the universe is driven by the difference between the degrees of freedom, $N_{sur}-N_{bulk},$ that is,
\begin{equation}
\label{loe_1}
\frac{dV}{dt}= L_P^2\left(N_{surf}-N_{bulk}\right).
\end{equation}
Here $V$ is the volume of the apparent horizon of the flat universe,  $L_p$ is the Planck length, in natural units $\hbar=1=C$ and hence $L_P^2=G$. $N_{sur}$ and $N_{bulk}$ are the degrees of freedom on the horizon and that within the bulk enclosed by the horizon, respectively. According to this, the universe is evolving towards a state that satisfies the holographic equipartition condition, $N_{sur}=N_{bulk},$  corresponds to the convention de Sitter epoch. From this law, Padmanabhan derived the Friedmann equations of a 3+1 dimensional flat universe in the context of Einstein's gravity. Cai has considered this law for an n+1 dimensional flat universe in Einstein, Gauss-Bonnet, and more general Lovelock gravity theories, by properly modifying the surface degrees of freedom and the volume increase \cite{Cai2012}. Later, The law of emergence has been extended for an n+1 dimensional non-flat universe with apparent horizon $\tilde{r}_{_A}$ as the boundary \cite{PhysRevD.87.061501},
\begin{equation}
\label{loe_2}
\alpha\frac{dV}{dt}=L_P^{n-1}\frac{\tilde{r}_{_A}}{H^{-1}}\left( N_{surf}-N_{bulk} \right),
\end{equation}
where $\alpha=\frac{n-1}{2(n-2)}$, $V=\Omega_n\tilde{r}_{_A}^n$ is the volume with $\Omega_n$ is the volume of unit n-sphere, and they derived the Friedmann equation for the universe with any spatial curvature. Here $\tilde{r}_{_A}$ is the radius of the apparent horizon, and $H$ is the Hubble parameter. The result has been extended to Gauss-Bonnet and Lovelock theories of gravity in the context of an n+1 dimensional universe. More generalizations for the law of emergence in different approaches are discussed in  \cite{PhysRevD.86.104013,PhysRevD.88.084019,FARAGALI2014335,Ai2014,PhysRevD.88.067303}. In exploring the connection with thermodynamics, the expansion law proposed 
in \cite{Cai2012,PhysRevD.87.061501}, 
has been derived from the unified first law of thermodynamics for Einstein, Gauss-Bonnet, and Lovelock gravity theories \cite{Mahith_2018}. Later, the motivation for choosing the volume as the areal volume in formulating the law of emergence has been explained and ratified from the thermodynamic point of view, is discussed in \cite{Hareesh_2019}. Recently, Pavon and Radicella assert that the entropy of the universe, with Hubble horizon as the boundary, evolves like that of an ordinary macroscopic system, approaching a state of maximum entropy \cite{Pav_n_2012}. Along this line, the authors \cite{Krishna:2017vmw} showed that the holographic equipartition is equivalent to a state of maximum entropy and the law of emergence describes a universe that evolves towards an end state, the de Sitter epoch of maximum entropy. The same authors extended this result to an n+1 dimensional Einstein, Gausss-Bonnet, and Lovelock gravity theories \cite{PhysRevD.99.023535,b2021emergence} as well. In the above works, the cosmological evolution in Einstein's gravity was studied using the Bekenstein entropy for the apparent horizon. However, recently, Barrow argued that quantum-gravitational effects might deform
the geometry of the black hole horizon, such that it can have an intricate fractal structure\cite{BARROW2020135643}. Accordingly, the area-entropy relation for the horizon gets modified as,
\begin{equation}
\label{entropy}
S = \left(\frac{A}{A_{0}}\right)^{1+\Delta/2},
\end{equation} 
which is later referred to as Barrow entropy. Here $A$ and $A_{0}$ are the area of the black hole horizon and Planck area, respectively, and  $\Delta,$ known as Barrow exponent, which quantifies the amount of quantum gravitational deformation effect on the black hole surface. The exponent constraint in the range $0\le\Delta\le1.$ For  $\Delta=0$, the Barrow entropy relation reduces to the Bekenstein entropy, and the most intricate fractal structure corresponds to the maximum value of $\Delta$. Although the expression for Barrow entropy resembles the non-extensive Tsallis entropy relation, the underlying principles in formulating these two are entirely different \cite{Tsallis2013,Dheepika2022}. By defining the Barrow holographic dark energy, the cosmological implications of Barrow entropy have been studied in reference \cite{PhysRevD.102.123525}. Since then, the cosmological aspects using Barrow entropy have become a potential area of research. Later, using Barrow holographic dark energy models with different length scales as infrared cut-off, the Barrow exponent $\Delta$ has been a constraint, by confronting with observational data \cite{Anagnostopoulos2020,Oliveros2022,doi:10.1142/S0218271822501073}. The generalized second law of thermodynamics is validated for the Barrow entropy associated with the horizon\cite{Saridakis2021}. From the idea of the connection between gravity and thermodynamics, the author in \cite{Saridakis_2020} obtained the modified Friedmann equations from thermodynamic relation $\mathcal{T}dS=-d\tilde{E}$ by using Barrow entropy instead of the usual Hawking-Bekenstein one. Meantime Sheykhi also obtained the dynamical equations from the unified first law of thermodynamics $dE=TdS+WdV$ and studied the entropy evolution in the model\cite{PhysRevD.103.123503}. Recent studies on Barrow entropy and its applications in the cosmological framework have been scrutinized in\cite{Mamon2021,BARROW2021136134,Sheykhi:2022jqq,Srivastava:2020cyk,Pradhan:2021cbj,Huang2021,PhysRevD.105.044042}. All these shows that Barrow entropy can replace the Bekenstein entropy in describing the evolution of the universe.

As we mentioned, an alternate description of cosmology is by using the law of emergence. In the present paper, our aim is to concentrate on this aspect, using Barrow entropy for the horizon instead of Bekenstein entropy. The first attempt along this direction is given in \cite{PhysRevD.103.123503}, where the author proposed a specific form of law of emergence by assuming an effective area $\tilde{A}=A^{1+\Delta/2},$ for the horizon having Barrow entropy as the horizon entropy. From which the modified Friedmann equation in 3+1 dimension was derived, is identical to the Friedmann equation obtained from thermodynamic principles\cite{PhysRevD.103.123503}. 
In the present work, we aim to derive the law of emergence from the thermodynamic laws by taking the entropy of the horizon as Barrow entropy. We have obtained the law of emergence for an n+1 dimensional universe from the unified first law and the Clausius relation. The result has been reduced to the corresponding 3+1 Einstein's gravity with $n=3.$ In contrast to the reference \cite{PhysRevD.103.123503}, we have first restructured the Barrow entropy into a convenient form, $S=\frac{A_{_{eff}}}{4G},$ which is similar in form to the Bekenstein entropy. Due to this, the law obtained in the present work is of relatively simple form.   \\

Recently, a turn occurred in the thermodynamic analysis when one uses non-standard entropies like Barrow entropy. The authors of \cite{Asghari2022}, while trying to derive the field equation from the equilibrium Clausius relation corresponding to Barrow entropy, came to know that in order to satisfy the energy-momentum conservation, the equilibrium Clausius relation should be replaced by an entropy balance relation contain an additional entropy change, which makes the thermodynamics of the system non-equilibrium. In fact, this has been pointed out earlier by Eling et.al\cite{PhysRevLett.96.121301}, that any curvature correction to the entropy demands a non-equilibrium thermodynamic treatment. Therefore, the Clausius relation is to be modified with an additional entropy production term $dS_{p}$, that $\delta Q/\mathcal{T}=dS+dS_{p}$. This additional entropy production term is determined by imposing the energy-momentum conservation  $\nabla^{\mu} T_{\mu \nu}=0$. Following this non-equilibrium thermodynamic analysis in the context of Barrow entropy, the authors \cite{Asghari2022} studied the $\sigma8$ tension and showed that the problem could be resolved to a certain extent. A similar analysis with Tsallis modified gravity is carried out in \cite{10.1093/mnras/stab2671} and claimed a slight alleviation of the $\sigma8$ tension. Following Eiling et al., many works have appeared in which a non-equilibrium thermodynamic approach has been adopted in gravity theories, which consists of higher-order curvature corrections. The authors in \cite{akbar2007thermodynamic} have shown that the Friedmann equation corresponding to the $f(R) $ gravity can be formulated from the non-equilibrium thermodynamic relation of the form  $ dE=TdS+WdV+TdS_{p}$. Likewise, such non-equilibrium thermodynamic perspectives are considered for scalar-tensor, $f(T)$ gravity theories \cite{AKBAR20067,cai2007unified,KazuharuBamba_2011}, to obtain the corresponding Friedmann equations. Similarly, the authors in \cite{Dezaki2015} obtained the law of emergence for a Kodama observer from non-equilibrium thermodynamic principles, the $f(R)$, and scalar-tensor gravity theories. While deriving the field equations when entropy associated with the horizon is of Barrow entropic form, Asghari and Sheykhi have shown that higher order curvature corrections arise, which in turn necessitates the use of non-equilibrium thermodynamic treatment.

Motivated by these works, we obtain the law of emergence from the modified first law of thermodynamics, which take account of the additional entropy production when one uses the Barrow entropy for the horizon. In this process, we found that the law of emergence in the context of non-equilibrium thermodynamics has a relatively simple form in terms of areal volume, in contrast to its form in the corresponding equilibrium perspective, where it adopts an effective volume to explain the emergence of cosmic space. Further, we have shown that the law of emergence in this context with Barrow entropy follows the generalized second law of thermodynamics and satisfies the entropy maximization condition so that the entropy is bounded at the end stage of evolution. 
\\ 

The paper is organized as follows, in section 2, the law of emergence is derived from the equilibrium thermodynamic relations for an n+1 dimensional non-flat universe. Section 3 addresses the necessity of the non-equilibrium treatment for non-standard entropies like Barrow entropy. This way, we restructured the form of entropy and derived the law of emergence from the non-equilibrium description of the unified first law and Clausius relation. Next section in 4. We checked the entropy maximization condition directly from the Barrow entropy relation and from the modified law of emergence obtained in both equilibrium and non-equilibrium perspectives. A comparative analysis of resulting laws and entropy maximization constraint relations are emphasized in the last section and conclude the results. Throughout the paper, we use the natural units $k_B$ = c = $\hbar$ = 1 for the sake of briefness.
\section{Law of emergence from the equilibrium thermodynamic principle }
This section, we will obtain the law of emergence for an n+1 dimensional non-flat universe from the unified first law of thermodynamics, with Barrow entropy for the horizon. Let us consider a homogeneous and isotropic n+1 dimensional Friedmann-Leima$\hat{i}$tre-Robertson-Walker (FLRW) universe with background metric given by, 
\begin{equation}
\label{m}
dS^2=h_{\mu\nu}dx^\mu dx^\nu+ \tilde{r}^2d\Omega^2.
\end{equation}
Where $\tilde{r}=a(t)r$ with $a(t)$ as the scale factor at time $t$, $\mu,$ and $\nu$ take values $0$ and $1$ such that, $(x^0, x^1) = (t, r)$ and $h_{\mu\nu}=diag[-1,a^2/1-kr^2]$ represents the two dimensional metric with the spatial curvature parameter $k,$ which can take the values $[-1, 0, +1]$ corresponding to closed, flat and open universe respectively. Also, $d\Omega^2=d\theta^2+\sin^2\theta d\phi^2$ and $(r,\theta,\phi)$ are the co-moving coordinates. The apparent horizon of the universe satisfies the relation $h^{\mu\nu}\partial_\mu\tilde{r} \partial_\nu\tilde{r}=0$, which in turn gives the expression for the radius of the apparent horizon as\cite{PhysRevD.49.6467,HAYWARD1999347}, 
\begin{equation}
\label{ra}
\tilde{r}_{_A}= \frac{1}{\sqrt{H^{2}+\frac{k}{a^{2}}}},
\end{equation} 
and its time rate is,
\begin{equation}
\label{eqn32.1}
\dot{\tilde{r}}_{_A}=-H\tilde{r}^{3}_{_A}(\dot{H}-\frac{k}{a^2}).
\end{equation}
Here $H=\dot{a}/a$ is the Hubble parameter, and the over dot represents the derivative with respect to cosmic time. For a flat universe, the apparent horizon becomes the Hubble horizon with radius, 
$\tilde{r}_{_A}=1/H$. The temperature associated with the apparent horizon is, $T=\kappa/2\pi$\cite{PhysRevD.75.084003}, where the surface gravity, $\kappa=\frac{1}{2\sqrt{-h}}\partial_a(\sqrt{-h}h^{ab}\partial_br)$, which takes the form\cite{Cai_2005,Dezaki:2014iaa},
\begin{equation}
\kappa=-\frac{1}{\tilde{r}_{_A}}\left(1-\frac{\dot{\tilde{r}}_{_A}}{2H\tilde{r}_{_A}}\right).
\end{equation}
Thus the horizon temperature becomes,
\begin{equation}
\label{T}
T=- \frac{1}{2\pi \tilde{r}_{_A}}
\left(1-\frac{\dot{\tilde{r}}_{_A}}{2H\tilde{r}_{_A}}\right).
\end{equation} 
But in the case of an expanding universe, $\frac{\dot{\tilde{r}}_{_A}}{2H\tilde{r}_{_A}} <1,$ \cite{Dezaki2015}, hence the temperature reduces to,
\begin{equation}
\label{clst}
\mathcal{T}=- \frac{1}{2\pi \tilde{r}_{_A}}.
\end{equation}
Hayward obtained the unified first law of equilibrium thermodynamics for a black hole
horizon as \cite{Hayward_1998,HAYWARD1999347},
\begin{equation}
\label{ufl}
 dE=A\psi+W dV,
\end{equation}
which can be applied to a cosmological apparent horizon \cite{DongsuBak2000,RongGenCai2005} as well.
Here, $E=\rho V$ is the Misner-Sharp energy contained within the apparent horizon of volume, $V=\Omega_n\tilde{r}^{n}_{_{_A}},$ with $\Omega_n=\frac{\pi^{n/2}}{\Gamma(\frac{n}{2}+1)},$ the volume of an n-sphere with unit radius, and $\rho$ is the density of the cosmic fluid enclosed by the horizon. Here the scalar invariant $W$ is the work density, defined by the relation $W=-\frac{1}{2} T^{ab} h_{ab},$ which for a universe with perfect fluid assumes the form,
\begin{equation}
\label{w}
W=\frac{1}{2}(\rho-P),
\end{equation}
where $P$ is the pressure of the fluid in the universe. The vector invariant, $\psi$ is interpreted as the energy flux density, which for a universe with perfect fluid can be expressed as, \cite{PhysRevD.90.104042},
\begin{equation}
\label{psi}
\psi=\psi^{t}+\psi^{\tilde{r}_{A}}=-(\rho+P)H\tilde{r}_{_A}dt+\frac{1}{2}(\rho+P)d\tilde{r}_{_A},
\end{equation}
where $\psi^{t}$ is the flux crossing the horizon per unit area during the infinitesimal interval of time $dt$, during which the horizon is stationary and $\psi^{\tilde{r}_{A}}$ is the flux density crossing the dynamic horizon. The heat energy $\delta Q,$ flowing through the horizon in an infinitesimal interval of time, $dt,$ during which the horizon is considered to be stationary, is given by the Clausius relation, $\delta Q = \mathcal{T} dS,$ where $\mathcal{T}$ is the temperature of the stationary horizon, with fixed radius and $dS$ is the entropy change. The corresponding energy change, $-d\tilde{E},$ inside the horizon satisfies, $\delta Q=-d\tilde{E}$. This energy change is, in turn, related to the flux density as $d\tilde{E}=A\psi^{t}.$ 
Hence we have,
\begin{equation}
 \label{cls}
 -d\tilde{E}=\mathcal{T}dS=-A\psi^{t}.
 \end{equation}
Taking account of these, the unified first law in equation (\ref{ufl}) can then be rewritten as,
 \begin{equation}
 \label{ufl2}
  dE=\frac{1}{2\pi \tilde{r}_{_A}}dS+A\psi^{\tilde{r}_{A}}+W dV.
 \end{equation}
The above equation can be multiplied throughout by  $\left(1-\dot{\tilde{r}}_{_A}/2H\tilde{r}_{_A}\right)$, and using conservation equation, it become,
 \begin{equation}
 dE=-TdS+\frac{n\Omega_n \tilde{r}_{_{A}}^{n-1}\dot{\tilde{r}}_{_A}dt}{2}({\rho+P}).
 \end{equation}
Here $T$ is the dynamic temperature and the above equation can be expressed for $dS$ as,
 \begin{equation}
\label{ufl3}
dS=-2\pi \Omega_n \tilde{r}_{_A}^{n+1}d\rho.
\end{equation} 
This equation gives the change in entropy of the horizon in relation to the size of the horizon and the change in the density of the cosmic fluid within the horizon. Now we assume
Barrow entropy for the horizon as given in equation (\ref{entropy}), and obtain the change of entropy $dS,$ on the left-hand side of the above equation. Accordingly, we obtained it as,
\begin{equation}
\label{entpchang}
 dS= \left(\frac{n\Omega_n}{A_{0}}\right)^{1+\Delta/2} (n-1)(1+\Delta/2)\quad \tilde{r}^{(n-1)(1+\Delta/2)-1}_{_A} d\tilde{r}_{_A}.
\end{equation}
Substituting this in to equation (\ref{ufl3}), along with $d\rho$ by using  conservation equation, $\dot{\rho}+nH(\rho+P)=0,$ we finally get, after a suitable integration, as  
\begin{equation}
\label{eqn4}
\tilde{r}^{\frac{(n-1)\Delta-4}{2}}_{_A}= \frac{16\pi G}{n(n-1)}\left(\frac{A_{0}}{n \Omega_n}\right)^{\Delta/2} \frac{\left(4-(n-1)\Delta\right)}{2+\Delta}\rho.
\end{equation}
Here we have set the integration constant to zero. It may be noted that, the above equation, in comparison with equation (\ref{ra}), is equivalent the Friedmann equation\cite{PhysRevD.103.123503}.
However, our aim is to obtain the law of emergence. Multiply the above equation with $a^2$ and differentiate with respect to cosmic time $t,$ the result can then be simplified using continuity equation, we arrive at, 
\begin{multline}
\label{eqn7}
 \tilde{r}^{(n-1)(1+\Delta/2)}_{_A} \, \dot{\tilde{r}}_{_A}= \left(\frac{2}{4-(n-1)\Delta}\right) \times \\H\tilde{r}_{_A} \left[2\tilde{r}^{(n-1)(1+\Delta/2)}_{_A}+\frac{2\pi\Omega_n}{(n-1)(2+\Delta)} \left(\frac{A_0}{n\Omega_n}\right)^{1+\Delta/2}\tilde{r}^{(n+1)}_{_A} \left(4-(n-1)\Delta\right)((n-2)\rho+nP)\right].
\end{multline}
To express the above equation into a form equivalent to the law of emergence, let us now restructure the Barrow entropy relation in equation (\ref{entropy}), into form, $S=\frac{A_{_{eff}}}{4G},$ similar to the standard Bekenstein entropy. Here the effective area $A_{_{eff}}$ will assumes the form,
\begin{equation}
\label{aeff}
A_{_{eff}}= \frac{(n\Omega_n)^{1+\Delta/2}}{A_0^{\Delta/2}} \tilde{r}^{(n-1)(1+\Delta/2)}_{_A}.
\end{equation}
Following this, the corresponding effective volume change can be obtained as \cite{Cai2012},
\begin{equation}
\label{eqn10}
\frac{dV_{_{eff}}}{dt}=\frac{\tilde{r}_{_A}}{n-1}\frac{dA_{_{eff}}}{dt}
= \left(\frac{2+\Delta}{2}\right) \frac{(n\Omega_n)^{1+\Delta/2}}{A_0^{\Delta/2}} \tilde{r}^{(n-1)(1+\Delta/2)}_{_A} \dot{\tilde{r}}_{_A}.
\end{equation}
In view of this, multiply equation (\ref{eqn7}) through out with a factor $\left(\frac{2+\Delta}{2}\right) \frac{(n\Omega_n)^{1+\Delta/2}}{A_0^{\Delta/2}},$ then left hand side of the resulting equation become the time rate of the effective volume. Taking account of this we modified the equation (\ref{eqn7}) as,
\begin{equation}
\label{eqn11}
\alpha \frac{dV_{_{eff}}}{dt}=\frac{G\tilde{r}_{_A}}{H^{-1}}\left[ \frac{2+\Delta}{4-(n-1)\Delta} \frac{n-1}{n-2} \frac{(n\Omega_n)^{1+\Delta/2}}{GA_0^{\Delta/2}} \tilde{r}^{(n-1)(1+\Delta/2)}_{_A}+\frac{4\pi\Omega_n}{n-2}\tilde{r}^{n+1}_{_A} \left((n-2)\rho+nP\right)   \right].
\end{equation}
The above equation is the law of emergence with Barrow entropy as the horizon entropy, for an n+1 dimensional non-flat universe with surface degrees of freedom $N_{surf}$ and bulk degrees of freedom $N_{bulk}$ are as follows,
\begin{equation}
\label{eqn12}
N_{surf}= \frac{2+\Delta}{4-(n-1)\Delta} \frac{n-1}{n-2} \frac{(n\Omega_n)^{1+\Delta/2}}{GA_0^{\Delta/2}} \tilde{r}^{(n-1)(1+\Delta/2)}_{_A}, \quad
N_{bulk}=-\frac{4\pi\Omega_n}{n-2}\tilde{r}^{(n+1)}_{_A} \left((n-2)\rho+nP\right).
\end{equation}
Thus the equation (\ref{eqn11}) acquires the conventional form of law of emergence as, 
\begin{equation}
\label{loeeffv}
\alpha \frac{dV_{_{eff}}}{dt}= \frac{G\tilde{r}_{_A}}{H^{-1}}\left(N_{surf}-N_{bulk}\right).
\end{equation}
Now we will show that, the same form  as given in the above equation can also be derived 
from the alternative thermodynamic principle, the Clausius equation, which connects the energy flux through the horizon and the corresponding entropy change.
Let us consider the Clausius relation as given in equation (\ref{cls}). With appropriate horizon temperature (\ref{clst}) along with $\psi^{t}$ from the flux density equation (\ref{psi}), this equation can be rewritten as,
\begin{equation}
\frac{1}{2\pi \tilde{r}_{_A}}dS=\Omega_n\tilde{r}^n_{_A} nH(\rho+P).
\end{equation}
On using the horizon entropy change in equation (\ref{entpchang}), the above equation become,
\begin{equation}
\label{eqn15.1}
\tilde{r}^{(n-1)(1+\Delta/2)-(n+2)}_{_A}d\tilde{r}_{_A}=- \frac{4\pi \Omega_n}{(n-1)(2+\Delta)} \left(\frac{A_{0}}{n \Omega_n}\right)^{1+\Delta/2}d\rho,
\end{equation}
on suitable integration, we get,
\begin{equation}
\label{eqn15.2}
\tilde{r}^{\frac{(n-1)\Delta-4}{2}}_{_A}= \frac{2\pi \Omega_n}{(n-1)(2+\Delta)} \left(\frac{A_{0}}{n \Omega_n}\right)^{1+\Delta/2}\left(4-(n-1)\Delta\right) \rho. 
\end{equation}
Further simplification using the same algebra for equation (\ref{eqn7}) leads to,
\begin{equation}
\label{eqn16}
\alpha \frac{dV_{_{eff}}}{dt}=\frac{G\tilde{r}_{_A}}{H^{-1}}\left[ \frac{2+\Delta}{4-(n-1)\Delta} \frac{n-1}{n-2} \frac{(n\Omega_n)^{1+\Delta/2}}{GA_0^{\Delta/2}} \tilde{r}^{(n-1)(1+\Delta/2)}_{_A}+\frac{4\pi\Omega_n}{n-2}\tilde{r}^{(n+1)}_{_A} \left((n-2)\rho+nP\right)   \right].
\end{equation}
With the suitable forms of $N_{bulk}$ and $N_{surf}$ this will become exactly similar to the form equation(\ref{loeeffv}).
It may be noted that, we obtained the law of emergence in terms of time rate of an effective volume (see equation (\ref{loeeffv})) instead of the areal volume, which will reduces to the conventional areal volume when $\Delta =0$ and the resulting law takes the form identical 
to that proposed by Sheykhi\cite{PhysRevD.87.061501} for an n+1 dimensional non-flat universe. Further for $n=3$ the law will in turn reduces to the original form proposed by Padmanabhan for a 3+1 dimensional flat universe.
\section{Law of emergence from the non-equilibrium thermodynamic principle}
In the previous section we obtain the law of emergence, by taking the Barrow entropy for the horizon, from an equilibrium thermodynamic point of view. To facilitate it, we re-arranged Barrow entropy relation in a form similar to Bekenstein entropy by defining an effective area, $A_{eff}.$ Assuming the validity of the conventional conservation law, we have obtained the law of emergence in terms of an effective volume $V_{eff},$ using the first law of thermodynamics. However, when one adopt non-Bekenstein type of entropy for the horizon, there arise issues with the validity of the equilibrium thermodynamic approach. Eiling, Guedens and Jacobson\cite{PhysRevLett.96.121301} have shown that, while one try to derive the field equation with non-Bekenstein type entropy for the horizon, the equilibrium clausius relation is no longer be valid due to the fact that the energy momentum conservation, $\nabla_\mu T^\mu_\nu=0,$ does not hold in such cases. They resolve this issue by replacing the conventional clausius relation with an entropy balance relation \cite{PhysRevLett.96.121301,book:18538,Asghari2022},  
\begin{equation}
\label{25.1}
\delta S=\frac{\delta Q}{\mathcal{T}}+dS_p.
\end{equation}
Where $dS_p$ is the corresponding additional entropy produced in the system due to the non-equilibrium thermodynamic process. This procedure has been adopted by Asghari and Sheykhi \cite{Asghari2022}, for a universe with Barrow entropy for the horizon, and derived the corresponding modified Einstein field equation(see equation.24 in \cite{Asghari2022}) as given below (in a re-arranged form),
\begin{equation}
\label{einstein}
G_{\mu \nu}=R_{\mu \nu}-\frac{1}{2}Rg_{\mu\nu}=\frac{8 \pi A^{1+\Delta/2}_0}{4A^{\Delta/2}} \left[T_{\mu\nu}-\frac{\Delta}{(2+\Delta)}T_{\mu\nu}+\frac{A_0^{-(1+\Delta/2)}}{2\pi}\left(\nabla_\mu\nabla_\nu A^{\Delta/2 } -\Box
A^{\Delta/2}g_{\mu \nu}\right)\right].
\end{equation} 
The above equation can be conveniently expressed in a more suitable form as,
\begin{equation}
\label{25.4}
G_{\mu \nu}= R_{\mu \nu}-\frac{1}{2}Rg_{\mu \nu}=8\pi G_{_{_{eff}}}T^{^{(eff)}}_{\mu \nu},
\end{equation}
where $G_{eff}$ is the effective gravitational coupling strength, 
\begin{equation}
\label{25.41}
G_{_{_{eff}}}=\frac{ A^{1+\Delta/2}_0}{4A^{\Delta/2}},
\end{equation}
and $T_{\mu\nu}^{^{(eff)}}$ is the effective energy-momentum tensor, which is defined as the entire term present inside the square bracket on the right hand side of equation (\ref{einstein}). The $T_{\mu\nu}^{^{(eff)}}$ 
comprises of the energy density and the pressure of all the physical matter present in the universe and also the curvature part that arises due to the modified gravity. Assuming that entire components in the effective energy-momentum tensor are of perfect fluid type, then it can be expressed in a metric independent form as,
\begin{equation}
\label{eqn25.5}
T^{\mu (eff)}_{\nu}=Diag[-\rho_{_{_{eff}}}, P_{_{_{eff}}}, P_{_{_{eff}}}, P_{_{_{eff}}}],
\end{equation}%
where $\rho_{_{_{eff}}}$ and $P_{_{_{eff}}}$ is the effective energy density and the corresponding pressure.
Following the Bianchi identities, the covariant derivative of the equation (\ref{25.4}) holds the generalized energy momentum conservation,  $ \nabla_\mu G^\mu_\nu=0=8\pi \nabla_\mu \left(G_{_{_{eff}}}T^{\mu (eff)}_{\nu}\right)$. This implies the generalized continuity equation as
\cite{PhysRevD.90.104042},
\begin{equation}
\label{con}
\dot{\rho}_{_{eff}}+nH(\rho_{_{eff}}+P_{_{eff}})=-\frac{\dot{G}_{_{eff}}}{G_{_{eff}}}\rho_{_{eff}}.
\end{equation} 
The right hand side of the above equation, $-\frac{\dot{G}_{_{eff}}}{G_{_{eff}}}\rho_{_{eff}}$ has the dimension of effective energy density flow, $ \dot{\rho}_{_{eff}}$ which balances the energy flow due to the non-equilibrium situation. The corresponding energy can then be defined as,
\begin{equation}
\label{epsi}
\mathcal{E}= -\Omega_n \tilde{r}^n_{_A}\frac{\dot{G_{_{eff}}}}{G_{_{eff}}}\rho_{_{eff}}dt,
\end{equation}
which can be taken as the energy dissipated in the system due to non-equilibrium thermal evolution.
Having the field equation and energy momentum tensor, one can obtain the Friedmann equations and for an n+1 dimensional universe with curvature parameter $k$ these equations take the form,
\begin{equation}
\label{feqn1}
\left(\dot{H}-\frac{k}{a^{2}}\right)=\frac{-8\pi G_{_{_{eff}}}}{(n-1)} ( \rho_{_{_{eff}}}+P_{_{_{eff}}}) \hspace{0.5cm};\hspace{0.5cm} \left(H^{2}+\frac{k}{a^{2}}\right)=\frac{16\pi G_{_{_{eff}}}}{n(n-1)}  \rho_{_{_{eff}}}.
\end{equation}
In reference \cite{PhysRevD.90.104042}, the above equations are obtained from the principles of thermodynamics, the Clausius rule and the unified first law, by assuming an effective entropy,
\begin{equation}
\label{eqn26}
\tilde{S}=\frac{A}{4G_{_{eff}}},
\end{equation}
with an effective coupling strength $G_{_{eff}}$. Such a definition of entropy is feasible due to form of the field equation (\ref{25.4}), in terms of $G_{_{eff}}$ as given in equation(\ref{25.41}). We adopt this definition of the entropy for deriving the law of emergence from the non-equilibrium modified unified law and also form the entropy balance relation for an n+1 dimensional universe with Barrow entropy for the horizon.
\subsection{From the corrected unified first law due to non-equilibrium behaviour}
In this section we are deriving the law of emergence from the unified first law of thermodynamics, which takes account of the energy dissipation due to the additional entropy production. Let us consider the law with an extra term, $\mathcal{E},$ which arise due to the energy dissipation of the non-equilibrium situation,
\begin{equation}
\label{uflw2}
dE_{_{_{eff}}}=A\psi_{_{eff}}+W_{_{_{eff}}}dV+\mathcal{E}
\end{equation}
Where $E_{_{_{eff}}}=\rho_{_{eff}} V$ with $\rho_{_{eff}}$ as an effective energy density and volume $V=\Omega_n\tilde{r}^{n}_{_A}$ with $\Omega_n$ being the volume of a n-sphere with unit radius, $dV=n\Omega_n\tilde{r}^{n-1}_{_A}d\tilde{r}_{_A}$. The effective flux energy density $\psi_{_{eff}},$ can be defined in terms of effective energy density and the corresponding pressure as,
\begin{equation}
\label{psi2}
\psi_{_{eff}}=\psi^{t}_{_{eff}}+\psi^{\tilde{r}_{A}}_{_{eff}} =-(\rho_{_{eff}}+P_{_{eff}})H\tilde{r}_{_A}dt+\frac{1}{2}(\rho_{_{eff}}+P_{_{eff}})d\tilde{r}_{_A},
\end{equation}
and the effective work energy density is, $W_{_{eff}}=(\rho_{_{eff}}-P_{_{eff}})/2$. Here $\psi^{t}_{_{eff}}$ and $\psi^{\tilde{r}_{A}}_{_{eff}}$ is the effective flux crossing the horizon per unit area during the infinitesimal interval of time $dt$, during which the horizon is stationary and the effective flux density crossing the dynamic horizon respectively. 
The energy dissipation within the system, causes an additional entropy production, and thus the Clausius relation will be modified as 
\cite{PhysRevD.90.104042} 
\begin{equation}
\label{eqn28}
\mathcal{T}d\tilde{S}+\mathcal{T}d\tilde{S}_p=-d\tilde{E}_{_{_{eff}}}.
\end{equation}
Here 
\begin{equation}
\label{Eeff}
d\tilde{E}_{_{_{eff}}}=A \psi^{t}_{_{eff}}+\mathcal{E},
\end{equation}
is the flux energy across the apparent horizon during the interval of time $dt$ 
\cite{PhysRevD.90.104042}.
We can then write 
\begin{equation}
\label{clausius}
\mathcal{T}d\tilde{S}+\mathcal{T}d\tilde{S}_p=-(A \psi^{t}_{_{eff}}+\mathcal{E}).
\end{equation}
The unified first law in (\ref{uflw2}), can then be rewritten as,
\begin{equation}
\label{ufl4}
dE_{_{_{eff}}}=-\mathcal{T}d\tilde{S}-\mathcal{T}d\tilde{S}_p+A\psi^{\tilde{r}_{A}}_{_{eff}}+W_{_{_{eff}}}dV.
\end{equation}
 Let us substitute for the  work energy density $W_{_{eff}}$ and $\psi^{\tilde{r}_{A}}_{_{eff}}$ in the above equation and multiply throughout by a factor $\left(1-\dot{\tilde{r}}_{_A}/2H\tilde{r}_{_A}\right)$, then the resulting equation(\ref{ufl4}) can be simplified by using the continuity equation yields an alternative form for unified first law due to non-equilibrium,
\begin{equation}
\label{altufl}
dE_{_{_{eff}}}=-T(d\tilde{S}+d\tilde{S}_p)+( \rho_{_{eff}}-P_{_{eff}})\frac{n\Omega_n \tilde{r}_{_A}^{n-1} \dot{\tilde{r}}_{_A}}{2}dt +\mathcal{E}\frac{\dot{\tilde{r}}_{_A}}{2H\tilde{r}_{_A}}.
\end{equation}	
Here $T$ is the dynamical temperature of the horizon as given in equation(\ref{T}). Substituting for $\mathcal{E}$ and after suitable rearrangements, we can express the entropy production term, $Td\tilde{S}_p$ due to the non-equilibrium behaviour as,
\begin{equation}
\label{tsp}
Td\tilde{S}_p=-Td\tilde{S}- \frac{n\Omega_n \tilde{r}_{_A}^{n-1}d\tilde{r}_{_A}}{2} \left[( \rho_{_{eff}}+P_{_{eff}}) +\frac{\rho_{_{eff}}}{nH}\frac{\dot{G}_{_{eff}}}{G_{_{eff}}}\right]+\Omega_n \tilde{r}_{_A}^{n}\left[nH(\rho_{_{eff}}+P_{_{eff}})+\frac{\dot{G}_{_{eff}}}{G_{_{eff}}}\rho_{_{eff}} \right]dt.
\end{equation} 
Now we substitute $d\tilde{S}$ in the above equation directly using the Barrow entropy (\ref{eqn26}), which is after feeding back the horizon area $A=n\Omega_n\tilde{r}^{n-1}_{_A}$ is,
\begin{equation}
\label{eqn45}
d\tilde{S}=\frac{n\Omega_n}{4}\left[\frac{(n-1)}{G_{_{eff}}}\tilde{r}^{n-2}_{_A}\dot{\tilde{r}}_{_A}-\tilde{r}^{n-1}_{_A}\frac{\dot{G}_{_{eff}}}{G^2_{_{eff}}}dt\right].
\end{equation}
Hence we have,
\begin{equation}
\label{dsp1}
Td\tilde{S}_p= \left(1-\frac{\dot{\tilde{r}}_{_A}}{2H\tilde{r}_{_A}}\right)\frac{n(n-1)\Omega_n \tilde{r}^{n-3}_{_A}}{4\pi  G_{_{eff}}}      \left[\dot{\tilde{r}}_{_A}+\frac{\tilde{r}_{_A}}{4}\left(1-\frac{2}{n-1}\right)\frac{\dot{G}_{_{eff}}}{G_{_{eff}}}\right]dt.
\end{equation}
On substituting the temperature $T$ from equation(\ref{T}), we get the rate of additional entropy as,
\begin{equation}
\label{dsp2}
\frac{d\tilde{S}_p}{dt}=-\frac{n(n-1)\Omega_n \tilde{r}^{n-2}_{_A}}{2 G_{_{eff}}} \left[\dot{\tilde{r}}_{_A}+\frac{\tilde{r}_{_A}}{4}\left(1-\frac{2}{n-1}\right)\frac{\dot{G}_{_{eff}}}{G_{_{eff}}}\right].
\end{equation} 
For n=3, the entropy production rate became,
\begin{equation}
\label{ep2}
\frac{d\tilde{S}_P}{dt}=-\frac{4\pi \tilde{r}_{_A}\dot{\tilde{r}}_{_A}}{G_{_{eff}}}.
\end{equation}
From the Friedmann equations (\ref{feqn1}) together with the conservation relation (\ref{con}), time rate of the radius of the apparent horizon is always positive, i.e.,
\begin{equation}
\label{rdot}
\dot{\tilde{r}}_{_A}=\frac{n}{2}H\tilde{r}_{_A}(1+\omega_{_{eff}}) \geq 0,
\end{equation}
where $\omega_{_{eff}}=P_{_{eff}}/\rho_{_{eff}}$ is the effective equation of state parameter, such that for an accelerating universe which evolves toward an end de Sitter epoch, it satisfies $\omega_{_{eff}}\ge-1.$
Hence, the additional entropy production due to non-equilibrium process behaves as a decreasing function of time.
Now combining the unified first law (\ref{altufl}) with equation (\ref{dsp1}), we get,
\begin{dmath}
\rho_{_{eff}}dV+Vd \rho_{_{eff}}=\frac{1}{2\pi \tilde{r}_{_A}}
\left(1-\frac{\dot{\tilde{r}}_{_A}}{2H\tilde{r}_{_A}}\right)\left\{\frac{n(n-1)\Omega_n \tilde{r}^{n-2}_{_A}d\tilde{r}_{_A}} {4G_{_{eff}}}-\frac{n\Omega_n \tilde{r}^{n-1}_{_A}}{4}\frac{\dot{G}_{_{eff}}}{G^2_{_{eff}}}-\frac{n(n-1)\Omega_n \tilde{r}^{n-2}_{_A}}{2 G_{_{eff}}} \left[\dot{\tilde{r}}_{_A}+\frac{\tilde{r}_{_A}}{4}\left(1-\frac{2}{n-1}\right)\frac{\dot{G}_{_{eff}}}{G_{_{eff}}}\right]\right\}dt+ \frac{(\rho_{_{eff}}-P_{_{eff}})}{2}dV-\Omega_n \tilde{r}^n_{_A}\frac{\dot{G_{_{eff}}}}{G_{_{eff}}}\rho_{_{eff}} \frac{\dot{\tilde{r}}_{_A}}{2H\tilde{r}_{_A}}dt.
\end{dmath}
This equation can be simplified to a form,
\begin{equation}
\frac{d\tilde{r}_{_A}}{\tilde{r}_{_A}^3}=\frac{8\pi G_{_{eff}}}{n-1}H(\rho_{_{eff}}+P_{_{eff}})dt.
\end{equation}
Further, using the continuity equation, $\rho_{_{eff}}+P_{_{eff}}=\frac{1}{nH}\left[-\frac{\dot{G}_{_{eff}}}{G_{_{eff}}}\rho_{_{eff}}-\dot{\rho}_{_{eff}}\right]$ the above equation can be written as,
\begin{equation}
-\frac{d\tilde{r}_{_A}}{\tilde{r}^3_{_A}}=\frac{8\pi}{n(n-1)}\frac{d}{dt}\left(\rho_{_{eff}}G_{_{eff}}\right).
\end{equation}
Which on integration gives,
\begin{equation}
\frac{1}{\tilde{r}^2_{_A}}=\frac{16 \pi}{n(n-1)}\rho_{_{eff}}G_{_{eff}},
\end{equation}
where we had neglected the integration constant.
Multiply with $a^2$ and differentiate with respect to time $t$, we get
\begin{equation}
2a\dot{a}\left(\frac{1}{\tilde{r}^2_{_{A}}}-\frac{\dot{\tilde{r}}_{_A}}{H\tilde{r}^3_{_A}}\right)=\frac{16 \pi}{n(n-1)}\left[2a\dot{a}G_{_{eff}}\rho_{_{eff}}+a^2 \dot{G}_{_{eff}}\rho_{_{eff}} +a^2 G_{_{eff}}\dot{\rho}_{_{eff}}\right].
\end{equation}
On multiplying with $H\tilde{r}^3_{_A}/2a\dot{a}$ and employing the relation  $\dot{G}_{_{eff}}\rho_{_{eff}}=-G_{_{eff}}\dot{\rho}_{_{eff}}-nHG_{_{eff}}(\rho_{_{eff}}+P_{_{eff}}) $, we get
\begin{equation}
-\dot{\tilde{r}}_{_A}=-H\tilde{r}_{_A}-\frac{16 \pi}{n(n-1)}nH\tilde{r}^3_{_A}G_{_{eff}}(\rho_{_{eff}}+P_{_{eff}})+\frac{16 \pi}{n(n-1)}H\tilde{r}^3_{_A}G_{_{eff}}\rho_{_{eff}}.
\end{equation}
This equation can be suitably rearranged as an expression for the rate of change of the volume of the horizon, by multiplying it through out with
$n\Omega_n\tilde{r}^{n-1}_A.$ Thus we arrive at,
\begin{equation}
\label{loe}
\alpha\frac{dV}{dt}=\frac{G_{_{eff}}\tilde{r}_{_A}}{H^{-1}}\left[\frac{n-1}{2(n-2)}\frac{n\Omega_n\tilde{r}^{n-1}_{_A}}{G_{_{eff}}}+\frac{4\pi \Omega_n\tilde{r}_{_A}^{n+1}}{n-2}\left(\left(n-2\right)\rho_{_{eff}}+nP_{_{eff}}\right)\right],
\end{equation}
with $\alpha=\frac{n-1}{2(n-2)}.$ On identifying and the number of degrees of freedom corresponding to the surface $N_{surf}$, and that of bulk $N_{bulk}$ as,
\begin{equation}
\label{N}
N_{surf}= \frac{n-1}{2(n-2)}\frac{n\Omega_n\tilde{r}^{n-1}_{_A}}{G_{_{eff}}} \hspace{0.5cm}; \hspace{0.5cm}N_{bulk}= -\frac{4\pi \Omega_n\tilde{r}_{_A}^{n+1}}{(n-2)}\left(\left(n-2\right)\rho_{_{eff}}+nP_{_{eff}}\right),
\end{equation}
respectively, the law of emergence takes the form,
\begin{equation}
\label{loeform2}
\alpha\frac{dV}{dt}=\frac{G_{_{eff}}\tilde{r}_{_A}}{H^{-1}}\left[N_{surf}-N_{bulk}\right].
\end{equation}
For n=3, $\alpha$ became unity and the corresponding number of surface degrees of freedom, $N_{surf}= 4\pi \tilde{r}^2_{_A}/G_{_{eff}} $ and that of bulk $ N_{bulk}= -(16\pi^2 \tilde{r}^4_{_A}/3)(\rho_{_{eff}}+3P_{_{eff}})$.
\subsection{From the Entropy balance relation due to the non-equilibrium behaviour}
In this section we derived the modified form of law of emergence from the clausius relation in non-equilibrium perspective. Let us now consider the clausius relation termed as the entropy balance relation which arises due to the non-equilibrium behaviour  in(\ref{eqn28}), which can expressed after substituting the energy flux change though the constant apparent horizon given in equation(\ref{Eeff}), as,
\begin{equation}
\label{clausius1}
\mathcal{T}d\tilde{S}+\mathcal{T}d\tilde{S}_p=-(A \psi^{t}_{_{eff}}+\mathcal{E})
\end{equation}
By substituting the Barrow entropy change (\ref{eqn45}), $\psi^{t}_{_{eff}}$ from equation(\ref{psi2}) and the energy dissipation from equation(\ref{epsi}), we get the time rate of additional entropy produced due to the non-equilibrium behaviour as,
\begin{equation}
\label{dsp2.1}
\frac{d\tilde{S}_p}{dt}=-\frac{n(n-1)\Omega_n \tilde{r}^{n-2}_{_A}}{2 G_{_{eff}}} \left[\dot{\tilde{r}}_{_A}+\frac{\tilde{r}_{_A}}{4}\left(1-\frac{2}{n-1}\right)\frac{\dot{G}_{_{eff}}}{G_{_{eff}}}\right]
\end{equation} 
It may be noted that, the above equation is similar to the additional entropy production rate obtained, in the previous section, (refer equation (\ref{dsp2})), using the unified first law. The clausius relation can now be written as,  
\begin{multline}
\frac{-n(n-1)\Omega_n}{8\pi G_{_{eff}}}\tilde{r}^{n-3}_{_{A}} \dot{\tilde{r}}_{_{A}}dt+\frac{n(n-1)\Omega_n}{16\pi}\tilde{r}^{n-2}_{_{A}} \frac{\dot{G}_{_{eff}}}{G^2_{_{eff}}}dt+\frac{n(n-1)\Omega_n}{4\pi G_{_{eff}}}\tilde{r}_{_A}^{n-3}d\tilde{r}_{_A}\\=n\Omega_n\tilde{r}^{n-1}(\rho_{_{eff}}+P_{_{eff}})H\tilde{r}_{_A}dt+\Omega_n\tilde{r}^n_{_A}\frac{\dot{G}_{_{eff}}}{G_{_{eff}}}\rho_{_{eff}}dt
\end{multline}
Which reduces to the simple form,
\begin{equation}
\frac{\dot{\tilde{r}}_A}{H\tilde{r}^3_{_A}}dt=-\frac{8\pi G_{_{eff}}}{n-1}\left[\rho_{_{eff}}+P_{_{eff}}\right]dt
\end{equation}
Substitute from the continuity equation, $\rho_{_{eff}}+P_{_{eff}}=\frac{1}{nH}\left[-\frac{\dot{G}_{_{eff}}}{G_{_{eff}}}\rho_{_{eff}}-\dot{\rho}_{_{eff}}\right]$ and integrate the resulting equation 
 (neglecting the integration constant), we get,
\begin{equation}
-\frac{d\tilde{r}_{_A}}{\tilde{r}^3_{_{A}}}=\frac{8\pi}{n(n-1)}\frac{d}{dt}\left(\rho_{_{eff}}G_{_{eff}}\right),
\end{equation}
which is also similar to that of equation  which we obtained in the previous section. Hence, following the same steps in the previous section, we get the same law of emergence as is equation(\ref{loeform2}) with same form the degrees of freedom as given equation (\ref{N}.)\\
It is to be noted that, even though the over-all forms of law of emergences in equilibrium and non-equilibrium perspectives, given in equations (\ref{loeeffv}) and (\ref{loeform2}) respectively, are similar, there exists much differences between them. The volume, the rate of change of which, appearing on the left hand side of the law, is turn out to be the effective volume in the case of equilibrium evolution, while it is the simple areal volume of the apparent horizon for non-equilibrium case. This is because of the restructured form of Barrow entropy in non-equilibrium scenario resembles the conventional Bekenstein entropy relation with the horizon normal area. More over there occurs substantial changes in the expressions for the degrees of freedoms between equilibrium and non-equilibrium situations, see equations (\ref{eqn12}) and (\ref{N}). In which the surface degrees of freedom obtained in non-equilibrium approach seems in more convenient form. An additional difference is the appearance of the gravitational constant, such that it appeared as an effective term, $G_{_{eff}}$ as defined in equation (\ref{25.41}) in the case of non-equilibrium thermodynamic situations. On analysing the result,
For the energy-momentum conservation is to be valid, we can argue that the universe with Barrow entropy as the horizon entropy preferably behaves as a non-equilibrium thermodynamic system with some additional entropy production.

\section{Maximization of horizon entropy}\label{smax}
In the cosmological context, based on the Hubble expansion history, the  universe behaves as an ordinary macroscopic system that evolves towards a maximum entropy state,  such that it satisfies  \cite{Pav_n_2012,book:2128746,PhysRevD.87.047302,PhysRevD.100.123545},
\begin{align*} 
\dot{S} \geq 0 \textrm{\hspace{0.5cm}for\hspace{0.1cm} always\hspace{0.5cm}} ;\hspace{0.5cm} \ddot{S}<0 \hspace{0.5cm} \textrm{for\hspace{0.1cm} at\hspace{0.1cm}least \hspace{0.1cm}later\hspace{0.1cm} time\hspace{0.1cm} of\hspace{0.1cm} evolution.}
\end{align*}
Where the over-dot represents the derivative with respect to cosmic time, here, the first inequality expresses the increase in entropy of the universe as it expands; that is the generalized second law of thermodynamics. The second one implies that the entropy function satisfies the convexity condition such that the entropy will be maximum at the end stage of the evolution. The total entropy comprises the entropy of all the cosmic components present inside the horizon and that of the horizon. Since the horizon entropy is much larger than that of the cosmic components present inside, the total entropy can be approximately taken to be equal to the entropy of the horizon \cite{Egan_2010}. Hence, here we restrict to analyse the evolution of the horizon entropy. We will first obtain the maximization condition directly using the Barrow entropy for the horizon. Alternatively, the same constraint relation will be obtained from the modified law of emergence derived in the context of both equilibrium and non-equilibrium thermodynamics. 
For this we adopt the approach described in reference  \cite{Krishna:2017vmw,PhysRevD.99.023535,b2021emergence}.
 
Now, taking Barrow entropy for the horizon  (\ref{entropy}), time rate of entropy (given in equation \ref{entpchang}) can be expressed as, 
\begin{equation}
\label{sdot}
\dot{S}= \left(\frac{n\Omega_n}{A_{0}}\right)^{1+\Delta/2} (n-1)(1+\Delta/2)\quad \tilde{r}^{(n-1)(1+\Delta/2)-1}_{_A}\dot{\tilde{r}}_{_{A}}.
\end{equation}
Since, $\dot{\tilde{r}}_A \geq 0 $ always (refer equation (\ref{rdot}) ), the constraint relation $\dot{S}\ge0$ holds during the entire evolution of the universe. Thus, the generalised second law of thermodynamics is satisfied.\\
To check the validity of the convexity condition, the second derivative of entropy can be obtained as,
\begin{equation}
\ddot{S}=\frac{n\Omega_n}{4G_{_{eff}}}(n-1)(1+\Delta/2)\tilde{r}_{_{A}}^{n-3}\left[\left((n-2)+(n-1)\Delta/2\right)\dot{\tilde{r}}_{_A}^2+\tilde{r}_{_{A}} \ddot{\tilde{r}}_{_{A}}\right].
\end{equation}
For the condition $\ddot{S}<0$ is to be valid, the quantity inside the square bracket on the right-hand side of the above equation must be less than zero, at least at the end stage of evolution. From equation (\ref{rdot}), the second derivative of the radius of the apparent horizon can take the form,
\begin{equation}
\label{rddot}
\ddot{\tilde{r}}_{_A}=\frac{n}{2}\tilde{r}_{_{A}}\left[\frac{n}{2}H^2(1+\omega_{_{eff}})^2+\dot{H}(1+\omega_{_{eff}})+H\dot{\omega}_{_{eff}}\right].
\end{equation}  
In this the effective equation of state $\omega_{_{eff}},$ will decreases as the universe expand. In the asymptotic limit, the universe evolves towards a de Sitter stage, for which $\omega_{_{eff}}\rightarrow-1$,
hence the terms containing $1+\omega_{_{eff}}$ goes to zero. Since $\omega_{_{eff}}$ is a decreasing function of cosmic time, $\dot{\omega}_{_{eff}}$ will be always negative. This implies that, at the end stage of the evolution, $\ddot{\tilde{r}}_{_A} <0$. This, in turn, implies that the convexity condition is satisfied if,
\begin{equation}
\label{c1}
\left| \left((n-2)+(n-1)\Delta/2\right)\dot{\tilde{r}}_{_A}^2 \right| < \left|\tilde{r}_{_{A}} \ddot{\tilde{r}}_{_{A }}\right|,
\end{equation}
is true, at least at the end stage of evolution. Since $\omega_{_{eff}}\rightarrow-1$ asymptotically for the later time, $\dot{\tilde{r}}_{_A}$ in the left side of the inequality relation goes to zero (refer equation  (\ref{rdot})). Hence the above relation holds asymptotically, which implies that the convexity condition is satisfied. Thus, the horizon entropy will be maximized at least at the end stage of evolution. For 3+1 dimension the above inequality will be reduced to,
\begin{equation}
|(1+\Delta)\dot{\tilde{r}}_{_{A}}^2|<|\tilde{r}_{_A}\ddot{\tilde{r}}_{_{A}}|.
\end{equation}
\subsection{Maximization condition from the law of emergence in equilibrium approach}
In this section, we have obtained the constraint relation from the modified law of emergence 
in the context of equilibrium thermodynamics. We consider the generalized form of modified law of emergence derived from the equilibrium thermodynamics is given in equation(\ref{loeeffv}),
and substitute the left hand side of that equation as, $\frac{dV_{_{eff}}}{dt}=\frac{4G\tilde{r}_{_{A}}}{n-1}\dot{S}.$  Here we have used equation (\ref{eqn10}), in which  $\dot{\tilde{r}}_{_A}$ is  written in terms of $\dot{S}$ (refer equation (\ref{entpchang})). Hence we get,
\begin{equation} 
\label{dots2}
\dot{S}=\frac{n-2}{2}H\left[N_{surf}-N_{bulk}\right],
\end{equation}
where $N_{surf}$ and $N_{bulk}$ is given in equation(\ref{eqn12}). In general, for an expanding universe, the rate of change of cosmic volume is always positive, so the holographic discrepancy, $\left[N_{surf}-N_{bulk}\right] \ge 0$ always. Substitute the degrees of freedom corresponding to surface and bulk from equation (\ref{eqn12}) into the above expression, then $\dot{S}$ becomes,
\begin{equation}
\dot{S}=\frac{(n-1)(2+\Delta)}{2}\left(\frac{n\Omega_n}{A_0}\right)^{(1+\Delta/2)}\tilde{r}_{_A}^{(n-1)\Delta/2+n-2}\dot{\tilde{r}}_{_A}.
\end{equation}
This is same as equation (\ref{sdot}) in the previous section. Since $\dot{\tilde{r}}_{_A}\ge0$ for an expanding universe, then the horizon entropy will never decrease and hence the generalised second law of thermodynamics is satisfied. 
To check the maximization condition, we find the second derivative of equation (\ref{dots2}), and is,
\begin{equation}
\label{eqn23}
\ddot{S}=\frac{n-2}{2}\dot{H}(N_{surf}-N_{bulk})+\frac{n-2}{2}H \frac{d}{dt}(N_{surf}-N_{bulk}).
\end{equation}
By substituting $N_{surf}$ and $N_{bulk}$, 
\begin{equation}
\ddot{S}=\frac{(n-1) (2+\Delta)}{2}\left(\frac{n\Omega_n}{A_0}\right)^{(1+\Delta/2)}\tilde{r}_{_A}^{(n-1)(1+\Delta/2)-2}\left[\left((n-2)+(n-1)\Delta/2\right)\dot{\tilde{r}}_{_A}^2+\tilde{r}_{_{A}} \ddot{\tilde{r}}_{_{A}}\right].
\end{equation}
It is to be noted that, in the asymptotic limit $\ddot{\tilde{r}}_{_{A}}< 0$ (refer equation (\ref{rddot})), hence the last product term inside the square bracket on the right hand side will be negative asymptotically. Hence the condition for the validity of the entropy maximization during the last stage of the universe is,
\begin{equation}
\label{c2}
\left|\left((n-2)+(n-1)\Delta/2\right)\dot{\tilde{r}}_{_A}^2\right| < \left|\tilde{r}_{_{A}} \ddot{\tilde{r}}_{_{A}}\right|,
\end{equation}
which is exactly similar to the constraint relation that we directly obtained from the Barrow entropy relation (refer the section \ref{smax}). 
\subsection{Maximization condition from the law of emergence in non-equilibrium approach}
Here we obtained the constraint relation for entropy maximization from the modified form of law of emergence due to non-equilibrium behaviour. We now adopt the same procedure as in the last section. 
Let us consider the generalized form of law of emergence obtained in the non-equilibrium scenario, (\ref{loeform2}) in which, $dV/dt= n\Omega_n\tilde{r}_A^{n-1}\dot{\tilde{r}}_A$, $V=\Omega_n\tilde{r}_A^{n}$ being the volume, ($\dot{\tilde{r}}_{_A}$ is from equation (\ref{entpchang})). We then have,
\begin{equation}
\label{nondots}
\dot{\tilde{S}}=\frac{(n-2)(2+\Delta)}{4}H\left[N_{surf}-N_{bulk}\right].
\end{equation}
By substituting $ N_{surf}$ and $N_{bulk}$ from equation (\ref{N}) along with equation the Friedmann equations (\ref{feqn1}), we then get,
\begin{equation}
\dot{\tilde{S}}=\frac{n\Omega_n}{4G_{_{eff}}}(n-1)(1+\Delta/2)\tilde{r}_{_A}^{n-2}\dot{\tilde{r}}_{_{A}}
\end{equation}
The above equation holds the condition $\dot{\tilde{S}}\ge 0$, since $\dot{\tilde{r}}_{_A}\ge0$, always,
implies that the generalised second law of thermodynamics is satisfied.
Now, in order to check the consistency of entropy maximization condition, the second derivative of equation (\ref{nondots}) is obtained, and is,
\begin{equation}
\ddot{\tilde{S}}=\frac{(n-2)(2+\Delta)}{4}\left[\dot{H}\left(N_{surf}-N_{bulk}\right)+H\frac{d}{dt}\left(N_{surf}-N_{bulk}\right)\right]
\end{equation}
On substituting the degrees of freedoms, we get,
\begin{equation}
\ddot{\tilde{S}}=\frac{n\Omega_n}{G_{_{eff}}}(n-1)(1+\Delta/2)\tilde{r}_{_A}^{n-3}\left[\left((n-2)+(n-1)\Delta/2\right)\dot{\tilde{r}}_{_A}^2+\tilde{r}_{_{A}} \ddot{\tilde{r}}_{_{A}}\right]
\end{equation}
To get the entropy to be maximized, $\ddot{\tilde{S}}<0$ in the long run, for which, as argued in the last subsection,  $\ddot{\tilde{r}}_{_{A}}<0$ at least for far future and which in turn implies, 
\begin{equation}
\label{c3}
\left|\left((n-2)+(n-1)\Delta/2\right)\dot{\tilde{r}}_{_A}^2\right| < \left|\tilde{r}_{_{A}} \ddot{\tilde{r}}_{_{A}}\right|,
\end{equation}
should be satisfied. This is exactly is similar to the constraint relation obtained from the equilibrium description. Since we have $\ddot{\tilde{r}}_{_{A}}<0$ and $\dot{\tilde{r}}_{_{A}}= 0$ at least for the later time of evolution (refer equations \ref{rdot} and \ref{rddot}), the negativity of $\ddot{\tilde{S}}$  is obvious at least for the later time of evolution and the entropy maximization condition is ensured. \\

It is to be noted from the previous discussion that, the constraint relations for entropy maximization obtained for the equilibrium, non-equilibrium perspective and directly from the Barrow entropy relation are the same. The reason for this identical nature of the constraints, especially between equilibrium and non-equilibrium situations, can be understood as follows. The rate of the production of additional entropy is negative (refer equations (\ref{dsp2}) and (\ref{dsp2.1})), and hence the generation of the additional entropy decreases as the universe evolves, and it finally approach a state of equilibrium. 

\section{Discussion and Conclusion}

The profound connection between gravity and thermodynamics led to the exciting speculations on gravity that, it could be an emergent phenomenon. Following this, Padmanabhan put forward the notion that the expansion of the universe can be considered as the emergence of cosmic space with cosmic time. Accordingly Padmanabhan proposed the law of emergence, which explains that cosmic expansion is driven by the holographic discrepancy. 
Later, the law of emergence was derived from the fundamental principles of thermodynamics and have shown that it implies the maximization of entropy at the end stage of the evolution. 
In these works the entropy of the apparent horizon is assumed to be of the Bekenstein entropy form. 

In the present work we have first derived  
the modified law of emergence for an (n+1) dimensional universe under equilibrium conditions, from the thermodynamic principles (the first law of thermodynamics and also the Clausius relation), by assuming Barrow entropy for the horizon. The derived law 
is different in its exact form 
from the corresponding form proposed (not derived) in a recent paper \cite{luciano2023emergence}.  For deriving the law 
we have rearrangement of the Barrow entropy in a more suitable form,
$S=A_{_{eff}}/4G$.  
For an appropriate limit, $\Delta=0$ and $n=3$, the resulting law
reduces to the conventional law in (3=1) dimension, in which the left-hand side contains the time rate of the areal volume within the apparent horizon. We have analysed the maximisation of the horizon entropy and obtained the constraint relation 
directly from the Barrow entropy relation and also from the modified law in the equilibrium condition and are found to be the same as expected. 
We have also analysed the generalised second law of thermodynamics and found that it is satisfied throughout the evolution of the universe.

Despite the analysis under equilibrium conditions, it is to be noted that, while deriving the Einstein field equation from the Clausius relation, with Barrow entropy for the horizon. it is found that the equilibrium Clausius relation is not valid due to the generation of additional entropy during the evolution. 
This demand a non-equilibrium treatment and hence
the equilibrium Clausius relation is replaced with an entropy balance relation, which incorporates the additional entropy. 
The corresponding modified field equation, by restructuring Barrow entropy $S=A/4G_{_{eff}}, $  analogues to the Bekenstein formula, leads to an effective energy-momentum tensor and an effective gravitational coupling strength. 
Following this 
we obtained the modified form of the law of emergence for an n+1 dimensional non-flat universe with the time rate of the areal volume in the left-hand side and the 
an effective coupling parameter along with the holographic discrepancy on the right hand side.

Comparing the resulting law of emergence in the non-equilibrium with that obtained for the equilibrium perspective, the following may be noted. Even though the overall appearance of the general form are alike, there seem to have many differences. In the law derived from the non-equilibrium thermodynamics, the surface degrees of freedom naturally arised as the conventional form proposed Sheykhi in an n+1 dimensional non-flat universe and reduced to the standard form proposed by Padmanabhan for a 3+1 dimensional flat universe. 
We checked the consistency of the modified law of emergence derived from the non-equilibrium description of the thermodynamic principle and found that the constraint relation corresponding to the maximisation condition is the same as that of the constraint obtained in the equilibrium approaches as well as the one extracted directly from the entropy relation. The reason for this similarity can be understood by analysing the form of additional entropy production, which makes the system out of equilibrium. That is, the additional entropy production rate decreases over time, and in the asymptotic limit, it will approach to zero. This guarantees that, the universe approaches equilibrium behaviour, at the end stage of the evolution. In summary, on treating the universe with Barrow entropy for the horizon, make it behaves as a thermodynamic system that is out of equilibrium, with an additional entropy production, but it evolves towards an equilibrium scenario at the end stage of evolution by maximising the entropy. 

\section*{Declaration of competing interest}

The authors declare that they have no known competing financial interests or personal relationships that could have appeared to influence the work reported in this paper.

\section*{Data availability}
No data was used for the research described in the article.

\section*{Acknowledgment}
We are thankful to  Dheepika M, Vishnu A Pai and Hassan Basari V T
for their valuable suggestions and discussions.
Nandhida Krishnan P is thankful to CUSAT and Government of Kerala for financial
assistance.


\end{document}